\documentclass[onecolumn,showpacs,amsmath,amssymb]{revtex4}
\usepackage{amsmath}
\usepackage{amssymb}
\usepackage{graphicx}% Include figure files
\usepackage{dcolumn}% Align table columns on decimal point
\usepackage{bm}% bold math
\usepackage{natbib}
\begin{document}

\title{Novel interface-selected waves and their influences on wave competitions}
%\shorttitle{Title} %Insert here a short version of the title if it exceeds 70 characters

\author{Xiaohua Cui$^1$, Xiaoqing Huang$^1$, Zhoujian Cao$^2$, Hong Zhang$^3$, and Gang Hu$^1$}\email{ganghu@bnu.edu.cn}

\affiliation{$^1$Department of physics, Beijing Normal University,
Beijing 100875, China}

\affiliation{$^2$Institute of Applied Mathematics, Academy of
Mathematics and System Science, Chinese Academy of Sciences, 55,
Zhongguancun Donglu, Beijing 100080, China}

\affiliation{$^3$Zhejiang Institute of Modern
Physics and Department of Physics, Zhejiang\\
University, Hangzhou 310027, China}

\date{\today}% It is always \today, today,

 %\pacs{47.54.-r}{Pattern selection; pattern formation}
 %\pacs{05.45.-a}{Nonlinear dynamics and chaos}
 %\pacs{04.30.Nk}{Wave propagation and interactions}

\begin{abstract}
The topic of interface effects in wave propagation has attracted
great attention due to their theoretical significance and practical
importance. In this paper we study nonlinear oscillatory systems
consisting of two media separated by an interface, and find a novel
phenomenon: interface can select a type of waves (ISWs). Under
certain well defined parameter condition, these waves propagate in
two different media with same frequency and same wave number; the
interface of two media is transparent to these waves. The frequency
and wave number of these interface-selected waves (ISWs) are
predicted explicitly. Varying parameters from this parameter set,
the wave numbers of two domains become different, and the difference
increases from zero continuously as the distance between the given
parameters and this parameter set increases from zero. It is found
that ISWs can play crucial roles in practical problems of wave
competitions, e.g., ISWs can suppress spirals and antispirals.
\end{abstract}

\maketitle

%\section{Section title}

%\section{\label{sec:level1}Introduction\protect}
The behaviors of waves around interface between two media have
attracted continual and great interest [1-7]. In linear optics we
are familiar with the problems of wave reflection and refraction,
which are predicted analytically. However, for nonlinear systems the
interface-related behaviors become much more complex and diverse,
and much less known.

The problems of wave propagation in linear and nonlinear media have
also attracted considerable attention. Recently, new observations of
inwardly propagating waves have stimulated considerable interest in
this field. For several centuries, scientists have known waves
propagating forwardly from wave source only, called normal waves
(NW). In recent years, different types of waves propagating toward
the wave source (called here antiwaves, AW) have been observed in
both linear optics [1, 8] and nonlinear oscillatory systems [9-14].
These new phenomena introduce completely new topics of the interface
problem. For instance, novel phenomenon of negative refraction has
been reported in both linear optics [1, 8] and nonlinear oscillatory
systems [3,7].

In the present paper, we find another completely new nonlinear
interface-phenomenon: interface of two different media can generate
waves, called here interface selected waves (ISWs). In a well
defined parameter surface the frequency and wave number (also wave
length) of ISWs are identical in two media with different
parameters, and they can be predicted analytically. Away but near
this surface ISWs still exist, though the above analytical
predictions are no longer available and wave numbers of ISWs in the
two domains are no longer identical. By varying parameter away from
this surface continuously, wave numbers change continuously, and the
difference of wave numbers in two domains also increases from zero
continuously. It is found that ISWs play crucial roles in practical
problems of wave competitions in oscillatory systems, e.g., in
suppressing spirals and antispirals.

We consider the following bidomain reaction-diffusion system

\begin{subequations}
\begin{align}
\frac{\partial \mathbf{U}_1}{\partial t}&=\mathbf{f}(\mathbf{U}_1,
\boldsymbol{\mu}_1, \boldsymbol{\nu}_1)+\mathbf{D}(\boldsymbol{\gamma}_1):\nabla^2\mathbf{U}_1 \\
\frac{\partial \mathbf{U}_2}{\partial t}&=\mathbf{f}(\mathbf{U}_2,
\boldsymbol{\mu}_2, \boldsymbol{\nu}_2)+\mathbf{D}(\boldsymbol{\gamma}_2):\nabla^2\mathbf{U}_2 \\
\mathbf{U}_i=(U_i^{(1)}&,\cdots, U_i^{(m)}), \quad
\boldsymbol{\mu}_i=(\mu _i^{(1)}, \cdots, \mu_i^{(p)}), \nonumber \\
\boldsymbol{\nu}_i=(\nu _i^{(1)}&,\cdots, \nu _i^{(q)}), \quad
\boldsymbol{\gamma}_i=(\gamma _i^{(1)}; \cdots, \gamma _i^{(l)}), \nonumber \\
&i=1, 2,\ \ m\geq2 \nonumber
\end{align}
\end{subequations}
where $\mathbf{D}(\boldsymbol{\gamma}_i)$ is a $m \times m$ matrix
with constant elements depending on $\boldsymbol{\gamma}_i$. The
function $\mathbf{f}$ and diffusion matrix $\mathbf{D}$ in the two
domains are identical because the same reaction-diffusion processes
occur in both sides. On the other hand, the dynamical evolutions in
each side may be different due to different control parameters
$(\boldsymbol{\mu}_1,\ \boldsymbol{\nu}_1,\ \boldsymbol{\gamma}_1)$
and $(\boldsymbol{\mu}_2,\ \boldsymbol{\nu}_2,\
\boldsymbol{\gamma}_2)$. We represent the interface between the two
domains by $I$, then the following boundary conditions are required
on $I$

\begin{equation}
\mathbf{U}_{1\{I\}}=\mathbf{U}_{2\{I\}},\ \frac{\partial
\mathbf{U}_1}{\partial n_{\{I\}}}=\frac{\partial
\mathbf{U}_2}{\partial n_{\{I\}}}
\end{equation}
where $\frac{\partial \mathbf{U}_i}{\partial n_{\{I\}}}$ indicates a
derivative of $\mathbf{U}_i$ over space variable along the direction
perpendicular to the interface $I$. We assume that
$\mathbf{U}_1=\mathbf{U}_2=0$ is a stable point of Eq.(1) and
$\boldsymbol{\mu}_i(\boldsymbol{\nu}_i)$ controls (not controls) the
linear terms of $\mathbf{U}_i$ for the reaction parts. Hopf
bifurcation with frequency $\omega_0$ is supposed to occur at
parameters
$\boldsymbol{\mu}_1=\boldsymbol{\mu}_2=\boldsymbol{\mu}_0$.
Moreover, we assume further that both $\boldsymbol{\mu}_1$ and
$\boldsymbol{\mu}_2$ are slightly beyond the Hopf bifurcation point

\begin{equation}
\sum_{j=1}^{p}({\mu_i}^{(j)} -{\mu_0}^{(j)})^2\ll 1,\ i=1,2
\end{equation}
At $\boldsymbol{\mu}_i$, $\mathbf{U}_i$ performs periodic
oscillation of frequency $\omega_i$, and $\omega_i$ approaches to
$\omega_0$ as $\boldsymbol{\mu}_i$ reduces to $\boldsymbol{\mu}_0$
($i=1,\ 2$). Under condition(3) Eq.(1) can be reduced to amplitude
equations, i.e., the bidomain complex Ginzburg-Landau equations
(BCGLE) by the approach standard for the derivative of single-domain
CGLE [15, 16].

\begin{subequations}
\label{eq:sec2ten}
\begin{equation}
\frac{\partial A_1}{\partial
t}=a_1(1-i\Omega_1)A_1-b_1(1+i\alpha_1)|A_1|^2A_1+c_1(1+i\beta_1)\nabla^2A_1
\end{equation}
\begin{equation}
\frac{\partial A_2}{\partial
t}=a_2(1-i\Omega_2)A_2-b_2(1+i\alpha_2)|A_2|^2A_2+c_2(1+i\beta_2)\nabla^2A_2
\end{equation}
\begin{equation}
A_{1\{I\}}=A_{2\{I\}}, \ \frac{\partial A_1}{\partial
n_{\{I\}}}=\frac{\partial A_2}{\partial n_{\{I\}}}
\end{equation}
\end{subequations}
where $(a_i,\ \Omega_i)$, $(b_i,\ \alpha_i)$ and $(c_i,\ \beta_i)$
are related to $\boldsymbol{\mu}_i-\boldsymbol{\mu}_0$,
$(\boldsymbol{\mu}_i-\boldsymbol{\mu}_0,\ \boldsymbol{\nu}_i)$ and
$(\boldsymbol{\mu}_i-\boldsymbol{\mu}_0,\ \boldsymbol{\gamma}_i)$,
respectively. The continuity conditions Eq.(4c) can be derived from
condition (2) because the transformation from $U_1$ to $A_1$ is
exactly the same as that from $U_2$ to $A_2$ (in the two domains,
amplitude equations are derived at a common Hopf bifurcation point
with an identical linear matrix at $\boldsymbol{\mu}_0$), and the
transformations from $U_i$ to $A_i$ ($i=1,2$) are determined only by
this linear matrix. In Eq.(4) $A_1$ and $A_2$ are complex variables
in each side of the interface. With scaling transformations we can
fix $a_1,\ b_1,\ c_1$ and $\Omega_1$, and the remaining 8 parameters
are irreducible for BCGLE systems. In the following study we will
set $a_1=b_1=c_1=1, \ \Omega_1=0$ for numerical simulations without
mentioning and all the theoretical formulas are given generally for
12 parameters. Without the interface interaction, the two media have
their single-domain planar wave solutions [2, 17]
\begin{subequations}
\label{eq:sec2ten}
\begin{equation}
A_i(x,t)=\sqrt{(\frac{1}{b_i}(a_i-c_ik_i^2))}e^{i(k_ix-\omega_i t)},
\ 0\leq k^2_i \leq \frac{a_i}{c_i}
\end{equation}
\begin{equation}
\omega_i=a_i(\alpha_i+\Omega_i)+c_i(\beta_i-\alpha_i)k^2, \ a_i,
b_i, c_i>0
\end{equation}
\end{subequations}
Waves in media $M_1$ and $M_2$ are classified to NWs and AWs under
the conditions [3, 12, 14]
\begin{subequations}
\label{eq:sec2ten}
\begin{eqnarray}
NWs: \ \omega_i a_i(\Omega_i+\alpha_i)<0 \ ; \ or, \ \omega_i
a_i(\Omega_i+\alpha_i)>0 \ and\ |\omega_i|>|a_i(\Omega_i+\alpha_i)|
\end{eqnarray}
\begin{eqnarray}
AWs: \ \omega_i a_i(\Omega_i+\alpha_i)>0 \ and\
|\omega_i|<|a_i(\Omega_i+\alpha_i)|
\end{eqnarray}
\end{subequations}
By AWs we mean waves with negative phase velocity, while both NWs
and AWs have positive group velocities [11, 12, 14]. Now we focus on
the interface related problems, and start from a one-dimensional
(1D) BCGLE system. We are interested in the problem how the
interface can significantly influence the system dynamics. In
Fig.1(a)-(c) we study the system evolutions at two different
parameter sets with random initial conditions, and find
characteristically different features in the asymptotic states. The
most significant and new observation is given in Figs.1(a) and 1(b)
where we find homogeneous planar waves moving in both media from
right to left with a constant velocity and transparently crossing
the interface. These homogeneous running waves originate from the
interface (see Fig.1(b)), therefore, called interface selected waves
(ISWs). The phenomenon of Fig.1(a) is surprising. With two media
having different control parameters we intuitively expect that the
waves in two media must have different wave numbers (even if both
sides may have the same frequency). This common feature is clearly
seen in Fig.1(c) where we observe uniform bulk oscillation in the
right medium and waves propagating from left to right in the left
medium, clearly manifesting the interface $I$. However, in Fig.1(a)
waves propagate seemly in a homogeneous medium without feeling any
difference of $M_1$ and $M_2$; the interface is transparent to the
waves. Moreover, the realization of these running waves is stable
against different initial perturbations. This characteristic
phenomenon can never exist in linear systems, and has never been
observed so far in nonlinear systems.

It is interesting that we can predict the frequency and wave number
of ISWs explicitly and exactly under certain parameter conditions.
For case of Fig.1(a) we can determine the frequency and wave number
by the following simple requirements:
\begin{eqnarray}
\omega_1(k_1) = \omega_2(k_2), \ k_1 = k_2, \ |A_1|=|A_2|
\end{eqnarray}
Inserting Eqs.(7) into Eq.(5) we obtain a unique set of solutions
$\omega_I, k_I$
\begin{subequations}
\label{eq:sec2ten}
\begin{equation}
k_{1}^2=k_{2}^2=k_{I}^2=\frac{a_1(\alpha_1+\Omega_1)-a_2(\alpha_2+\Omega_2)}{c_2(\beta_2-\alpha_2)-c_1(\beta_1-\alpha_1)}
\end{equation}
\begin{equation}
\omega_1=\omega_2=\omega_I=\frac{a_1(\alpha_1+\Omega_1)c_2(\beta_2-\alpha_2)-c_1(\beta_1-\alpha_1)a_2(\alpha_2+\Omega_2)
}{c_2(\beta_2-\alpha_2)-c_1(\beta_1-\alpha_1)}
\end{equation}
And these solutions are exact in the parameter surface
\begin{gather}
for\ c_1b_2-c_2b_1 \neq 0:\ if\ and\ only\ if\
\frac{a_1b_2-a_2b_1}{c_1b_2-c_2b_1}-\frac{a_1(\alpha_1+\Omega_1)-a_2(\alpha_2+\Omega_2)}{c_2(\beta_2-\alpha_2)-c_1(\beta_1-\alpha_1)}=0\\
\nonumber for\ c_1b_2-c_2b_1=0:\ if\ and\ only\ if\ a_1b_2-a_2b_1=0
\end{gather}
\end{subequations}
which is obtained by inserting Eq.(8a) into Eq.(5a), and by the
identifying $|A_1|=|A_2|$. It can be easily confirmed that on the
parameter surface Eq.(8c) the planar wave solution Eq.(5) with
frequency and wave numbers given by Eqs.(8a) and (8b) are exact
solution of BCGLE. Moreover, the predictions of Eqs.(7) and (8)
agree exactly with the numerical results of Fig.1(a).

In Figs.2(a) and 2(b) we specify the surface of Eq.(8c) in some
parameter planes where solid lines represent the parameters
satisfying condition (8c). In Figs.2(c) and 2(d) we vary parameters
along the solid line of Fig.2(a) and numerically compute 1D BCGLE.
we compare the numerical results (empty cycles and triangles) with
theoretical predictions of Eqs.(8a) and (8c) (solid lines), and
find: (i) ISWs exist in a large area of surface Eq.(8c); (ii) the
predictions of Eqs.(8a) and (8b) coincide with numerical results
exactly (within computation precision). In Fig.2(e) we fix a
parameter set on the surface (black disk T) and present the
asymptotic pattern evolution of the BCGLE system. It is clearly
shown that ISWs, with the wave number and frequency predicted by
Eqs.(8a) and (8b), asymptotically control the entire bidomain during
their propagation. For demonstrating the possibility of observation
of ISWs in experiments, we study a reaction-diffusion model:
bidomain Brusselator. In Fig.2(f), we show ISWs of this chemical
model, satisfying all conditions of Eqs.(7).

The solutions of Eqs.(8a) and (8b) are exact for BCGLE only on the
parameter surface of Eq.(8c). Slightly away from this surface
Eqs.(8a) and (8b) can no longer predict the wave numbers and the
frequencies of ISWs exactly. In Figs.3(a) and 3(b) we compare the
theoretical predictions of Eqs.(8a) and (8b) with numerical results
of frequencies and wave numbers by varying parameters along the
dashed line in Fig.2(a). We find: (i) The solutions of Eqs.(8a) and
(8b) are not exact; the wave numbers in the two domains deviate from
each other (about $7.7\%$ difference in Fig.3(d)); (ii) However,
slightly away from the surface Eq.(8c), the feature that the
interface generates waves is still clearly observed (compare
Fig.1(b) with Fig.3(c)), i.e., ISWs still exist; (iii) By
continuously increasing the parameter distance from condition
Eq.(8c), the deviation of the numerical results from the theoretical
predictions Eqs.(8a) and (8b) increases continuously from zero too,
and for small parameter deviation the solutions Eqs.(8a) and (8b)
can still be used for predicting frequency and wave numbers with
very good approximation. In Figs.3(c) and 3(d) we show ISWs for a
parameter set away from the surface Eq.(8c) (Disk Q in Fig.2(a)). It
is clear that even away from surface Eq.(8c) the waves of Fig.3(c)
are generated by the interface in the similar way as in Fig.1(b)
though we have $k_1=k_2$ in Fig.1(b) but $k_1\neq k_2$ in Fig.3(c).
Thus the waves in Fig.1(a), 1(b) and Figs.3(c), 3(d), have obviously
the same interface-selected nature which is essentially different
from the waves of Fig.1(c). Similar ISWs with $k_1 \neq k_2$ can
also be observed in bidomain Brusselator. In Figs.3(e) and 3(f) we
take parameter set far away from that in Fig.2(f), and can still
observe ISWs. Here the wave numbers in the two sides have slight
difference ($\frac{2|k_1-k_2|}{|k_1+k_2|}\approx 6.77\%$).

There are some necessary conditions for ISWs to appear. Let us
analytically specify some of these conditions under Eq.(8c)
(Eqs.(8a) and (8b) are exact solutions of $\omega_I$ and $k_I$).
From Eqs.(8a) and (5a) we have an obvious necessary existence
condition for ISWs, i.e.,
\begin{subequations}
\begin{equation}
0 \leq
k_{I}^2=\frac{a_1(\alpha_1+\Omega_1)-a_2(\alpha_2+\Omega_2)}{c_2(\beta_2-\alpha_2)-c_1(\beta_1-\alpha_1)}
\leq \frac{a_i}{c_i}, \ i=1,2
\end{equation}
which should be satisfied because wave number $k_I$ must be real. If
this condition is violated, there is no physically meaningful
solutions of $k_I$ and $\omega_I$, and thus no ISWs can be observed.
This is the case of Fig.1(c). ISWs are generated by the interface
and propagate along a certain direction. Therefore, the waves must
forwardly propagate in one domain. Precisely, ISWs are NWs in the
left (or right) domain while AWs in the right (left) domain for
waves propagating from right (left) to left (right), and this
requires another parameter condition
\begin{gather}
AWs:\ \omega_I a_i(\Omega_i+\alpha_i)>0\ and\
|\omega_I|<|a_i(\Omega_i+\alpha_i)|\ \\ \nonumber NWs:\ \omega_I
a_{\bar{i}}(\Omega_{\bar{i}}+\alpha_{\bar{i}})<0\ or\ \omega_I
a_{\bar{i}}(\Omega_{\bar{i}}+\alpha_{\bar{i}})>0\ and\
|\omega_I|>|a_{\bar{i}}(\Omega_{\bar{i}}+\alpha_{\bar{i}})|\
\\ \nonumber i=1,\ \bar{i}=2\ or\ i=2,\ \bar{i}=1
\end{gather}
\end{subequations}
In order to provide an idea how these conditions influence the
existence of ISWs, we show Fig.4, where one can numerically observe
ISWs in the regions enclosed by disks, called "ISW" regions. In
Fig.4 dashed-dotted lines with $k^2_I=0$ and $\alpha_i=\beta_i,\
i=1\ or\ 2$ are the boundaries of "ISW" theoretically predicted by
Eqs.(9a)and (9b), respectively. In "No ISW" regions ISWs do not
exist due to violations of conditions of Eq.(9a) or (9b). In the
regions "Unstable", ISWs exist while waves with wave number $k_I$
are unstable due to Eckhaus instability, and there ISWs cannot be
numerically observed. Figures 4(a) and 4(b) are plotted in a small
parameter surface under the condition of Eq.(8c). Similar structure
of distributions "ISW", "No ISW" and "Unstable" regions can be
observed when parameters are varied slightly away from the set of
Eq.(8c).

Though the above investigations are made for 1D bidomain systems,
the observations of ISWs exist generally for high-dimensional
systems. In 2D oscillatory systems, much richer types of waves,
including spirals and antispirals, can be self-sustained, and wave
competitions become an important issue. Now we explore how ISWs play
crucial roles in wave competitions. We consider a 2D BCGLE system
with an interface line $II'$ in between. Without the interface
interaction, $M_1$ supports normal spirals (Figs.5(a), (d), (g)) and
$M_2$ supports antispirals ( Figs.5(b), (e), (h)). With the
interface interaction we find characteristically different results
of wave competitions. Fig.5(c): the antispiral wins the competition
and dominates the system with frequency $\omega_2$; Fig.5(f): the
spiral wins; Fig.5(i): ISWs win and dominate the two domains. The
reasons why we can observe so diverse results in similar
competitions between spiral and antispiral waves can be completely
understood based on the analysis of ISWs.

For explaining the results of Fig.5 we briefly introduce some known
conclusions on wave competitions in oscillatory systems. If
competitions occur in a homogeneous medium, the results are [3, 18]:
\begin{subequations}
\label{eq:sec2ten}
\begin{equation}
NWs \ against \ NWs: \ faster \ waves \ win
\end{equation}
\begin{equation}
AWs \ against \ AWs: \ slower \  waves \ win
\end{equation}
\begin{equation}
NWs \ against \ AWs: \ NWs \ win
\end{equation}
\end{subequations}
With the competition rules Eq.(10) and the analytical results of
Eqs.(7)-(9) we can fully understand and predict the diverse results
of Fig.5.

Inserting the parameters of Fig.5(c) into Eq.(8) we have
$\omega_I=0.636$. Considering conditions Eq.(6) we conclude that
ISWs are NWs in $M_1$ and AWs in $M_2$ (note, the interface is the
source of ISWs). The frequencies of the spiral in $M_1$ and the
antispiral in $M_2$ are $\omega_1=0.288$ and $\omega_2=0.626$,
respectively. According to conclusion (10) ISWs win the competition
in $M_1$ against the spiral while losing the battle in $M_2$ against
the AW spiral. Therefore, the antispiral waves of frequency
$\omega_2$ finally dominate the whole system. The parameters of
Fig.5(f) do not satisfy condition Eq.(9a), and no ISWs can be
generated. In Figs.5(d) and (e) we observe $\omega_1=-0.0567$ and
$\omega_2=0.151$. According to Eq.(6) waves of $\omega_1(\omega_2)$
are NWs (AWs) in both $M_1$ and $M_2$. On the basis of (10c), the
spiral of frequency $\omega_1$ wins the competition. The most
interesting observation is given in Fig.5(i) where we have
$\omega_I=0.932$. The frequency of the spiral(antispiral) in $M_1$
($M_2$) is $\omega_1=0.293$ ($\omega_2=1.037$). Therefore, ISWs win
both competitions against the spiral in $M_1$ (condition (10a)) and
against the antispiral in $M_2$ (condition (10b)). The asymptotic
state is ISWs in 2D system where ISWs suppress both spiral and
antispiral in Fig.5(i).

In conclusion, we investigated the role played by interfaces. A new
type of waves, interface selected waves (ISWs) were found in
bidomain systems with one medium supports AWs and the other NWs.
When control parameters are on a well defined parameter surface ISWs
propagate with analytically predictable same frequency and wave
number in two media with different parameters. When the parameters
are away but near this surface, ISWs can be also observed, of which
the frequency and wave numbers can be located approximately. These
waves are selected by interfaces between two media, and some
necessary conditions for observing these ISWs are specified. These
ISWs play important roles in wave competitions. For instance, under
certain conditions ISWs can suppress spiral and antispiral waves in
both media. These roles are important in practical applications.
Experimental realizations of ISWs in chemical reaction-diffusion
systems are strongly suggested, based on the well-behaved ISWs of
Fig.2(f), Fig.3(e) and Fig.3(f) computed for a chemical
reaction-diffusion model.

\acknowledgments

Z. Cao is supported by the knowledge innovation program of the
Chinese Academy of Sciences.

\bibliography{apssamp}

%\newpage

\begin{figure}[h]
\includegraphics[scale=0.3]{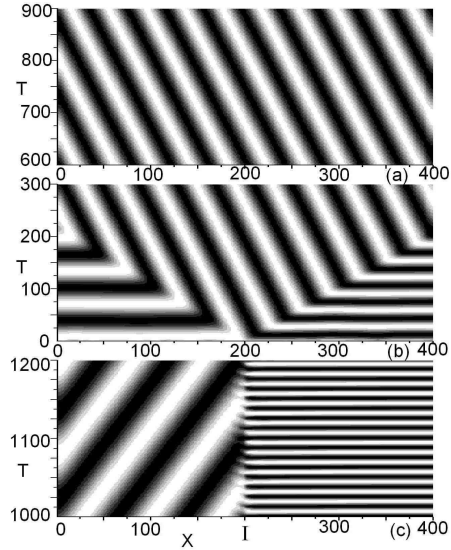}\\

\caption{(a)-(c) Contour patterns of Re$A_i$ of a 1D BCGLE system
with interface $I$. Domains $M_i,\ (i=1,2)$ have length $L$ and
parameters $\alpha_i,\ \beta_i$, and $a_i=b_i=c_i=1,\ \Omega_i=0,\
i=1,2$. Numerical simulations are made with space step $dx=0.5$,
time step $dt=0.005$, and $L=200$. No-flux boundary condition,
randomly chosen initial conditions and the above time and space
steps are used in all figures for numerical simulations unless
specified otherwise. (a) $\alpha_1=0.1,\ \beta_1=-1.4,\
\alpha_2=-0.2,\ \beta_2=1.4$. Interface-selected waves (ISWs)
homogeneous in $M_1$ and $M_2$ are observed, and the interface is
transparent to ISWs. (b) The same as (a) with early time evolution
plotted. It is clearly shown that ISWs originate from the interface.
(c) $\alpha_1=0.1,\ \beta_1=-1.4,\ \alpha_2=0.5,\ \beta_2=1.4$.
$M_1$ and $M_2$ support different regular waves, clearly manifesting
interface $I$.} \label{fig.1}
\end{figure}

\begin{figure}[h]
\includegraphics[scale=0.4]{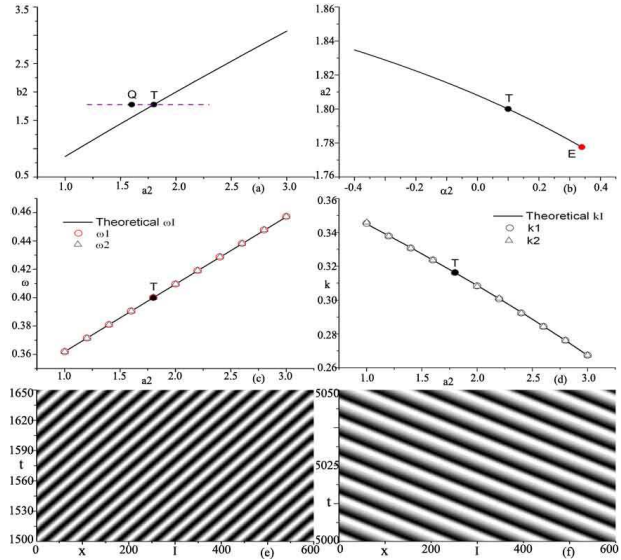}\\

\caption{(a) Surface Eq.(8c) (Solid line) is plotted in $a_2-b_2$
plane with $\Omega_2=0,\ c_2=2.0,\ \alpha_1=0.6,\ \beta_1=-1.4,\
\alpha_2=0.1,\ \beta_2=1.2$. (b) The same as (a) with surface (8c)
plotted in $\alpha_2-a_2$ plane, $b_2=1.778$. Point E corresponds to
the boundary $k^2=0$. (c) Frequency of ISWs which are selected along
the solid line of (a). Theoretical prediction Eq.(8a) (solid line)
coincides with numerical simulation for 1D BCGLE (empty circles and
triangles) perfectly. (d) The same as (c) with wave numbers
$k_1=k_2$ plotted. Agreement between theoretical prediction Eq.(8a)
and numerical results is also confirmed. (e) ISW pattern obtained by
using the parameter set $a_2=1.8,\ b_2=1.778$ (point T in (a)). (f)
The same as (e) with contour pattern of $v$ variable of Brusselator
RD which is numerically computed for 1D chain. The system is:
$\frac{\partial u_i}{\partial
t}=a_i-(b_i+1+\gamma_{i})u+(1+\sigma_{i})u^2v+\delta_{ui}\nabla^2u,\
\frac{\partial v_i}{\partial
t}=b_iu-(1+\sigma_{i})u^2v+\delta_{vi}\nabla^2v,\ i=1,2,\
 a_1=1.0,\ b_1=2.24,\ \gamma_{1}=\sigma_{1}=0.0,\ \delta_{u1}=2.31,\ \delta_{v1}=2.17;\ a_2=1.02,\ b_2=2.2624,\
 \gamma_{2}=0.02,\ \sigma_{2}=0.01,\ \delta_{u2}=0.95,\ \delta_{v2}=2.47;\ dx=dy=0.5,\ dt=0.0025,\ L=300$. ISWs with identical $w$ and $k$ are observed.}
\end{figure}

\begin{figure}[h]
\includegraphics[scale=0.4]{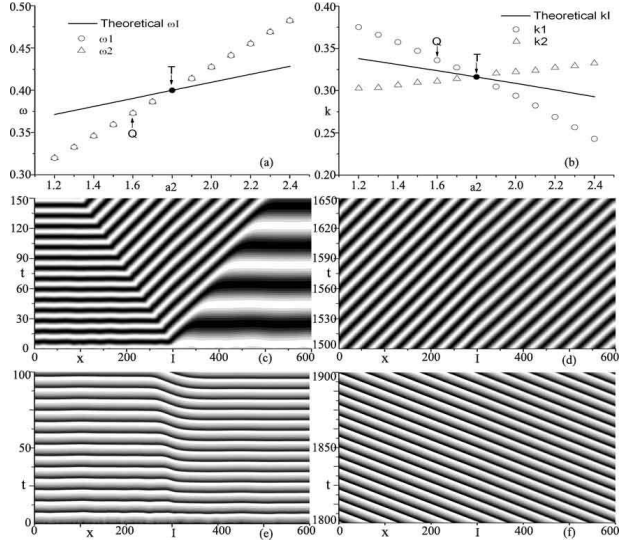}\\

\caption{(a) (b) The same as Figs.2(c) and (d), respectively, with
 parameters varied along the dashed line of Fig.2(a). Numerical simulations are made for 1D BCGLE. Now deviations
between theoretical results of Eqs.(8a), (8b) and the numerical
results are observed. Deviation increases as parameters vary away
from the surface Eq.(8c). (c) (d) The same as Fig.2(e) with
different time intervals plotted, respectively, in which parameters
are taken away from the solid line of Fig.2(a) (point Q,\ $a_2=1.6,\
b_2=1.778$). Now ISWs are still observed while wave numbers in the
two sides are slightly different($\frac{2|\triangle
k|}{|k_1+k_2|}\approx 7.71\%,\ \triangle k=|k_1-k_2|$). (e) (f) The
same as (c) and (d), respectively, with 1D Brusselator chain
computed. The parameter set is considerable different from that of
Fig.2(f): $a_1=1.0,\ b_1=3.2,\ \gamma_{1}=\sigma_{1}=0.0,\
\delta_{u1}=1.0,\ \delta_{v1}=0.5;\ a_2=1.1,\ b_2=3.2,\
 \gamma_{2}=0.1,\ \sigma_{2}=0.0,\ \delta_{u2}=1.0,\ \delta_{v2}=2.5$. Now the
wave numbers of the two sides are also slightly different
($\frac{2|\triangle k|}{|k_1+k_2|}\approx 6.77\%$).} \label{fig.3}
\end{figure}

\begin{figure}[h]
\includegraphics[scale=0.4]{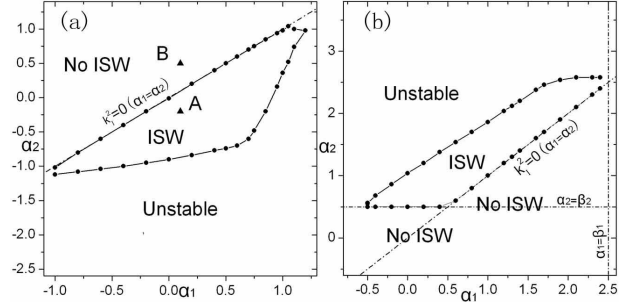}\\

\caption{The distributions of different types of waves in
($\alpha_1, \ \alpha_2$) parameter planes for different sets of
($\beta_1, \ \beta_2$). Black disks represent the boundaries of ISW
regions ("ISW" regions) identified by direct numerical simulations
of 1D BCGLE. In "No ISW" regions, ISWs do not exist due to the
violations of condition Eq.(9a) (boundary $k^2_I=0$, i.e.,
$\alpha_1=\alpha_2$) or condition Eq.(9b) (boundary
$\alpha_i=\beta_i,\ i=1$ or $2$) (both presented in dashed-dotted
lines). In the region "Unstable", both conditions Eqs.(9a) and (9b)
are satisfied, but waves with the given $k_I$ are unstable due to
the Eckhaus instability. $a_i=b_i=c_i=1,\ \Omega_i=0,\ i=1,2.$ (a)
$\beta_1=-1.4, \beta_2=1.4$, (b) $\beta_1=2.5, \beta_2=0.5$. Black
triangles A and B in Fig.4(a) represent the parameter sets used in
Figs.1(a), (c), respectively.}
\end{figure}

\begin{figure}[h]
\includegraphics[scale=0.3]{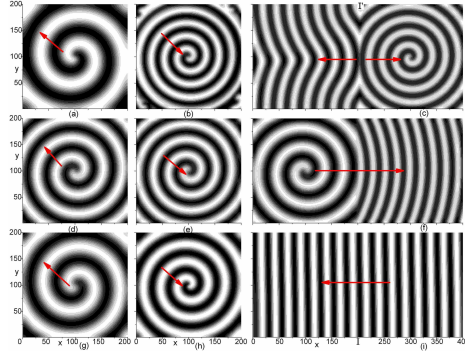}\\

\caption{Wave competitions between spiral, antispiral and ISWs in 2D
BCGLE with interface $II'$. $a_i=b_i=c_i=1,\ \Omega_i=0,\ i=1,2$.
(a)(d)(g) Spirals in $M_1$ medium. (b)(e)(h) Antispirals in $M_2$
medium. Snapshots in (a)(b), (d)(e) and (g)(h) are used as the
initial conditions for the dynamic evolutions of (c), (f) and (i),
respectively. (c)(f)(i) The asymptotic states of bidomain systems
with interface $II'$. (a)(b)(c) $\alpha_1=0.2,\ \beta_1=2.0,\
\alpha_2=1.0,\ \beta_2=-0.5$. Antispiral initially in $M_2$
dominates the system in (c). (d)(e)(f) $\alpha_1=0.2,\
\beta_1=-2.0,\ \alpha_2=0.5,\ \beta_2=-1.4$. Spiral initially in
$M_1$ dominates the system in (f). (g)(h)(i) $\alpha_1=0.1,\
\beta_1=3.2,\ \alpha_2=1.2,\ \beta_2=0.0$. In (i) ISWs suppress both
spiral in $M_1$ and antispiral in $M_2$, and dominate the whole
system.} \label{fig.4}
\end{figure}

\end{document}